# $X_1(2900)$ as a $\bar{D}_1 K$ molecule

Hao Chen[1,a], Hong-Rong Qi[2,b], Han-Qing Zheng[3,c]

[1] Department of Physics and State Key Laboratory of Nuclear Physics and Technology, Peking University, Beijing 100871, China
[2] Department of Engineering Physics, Tsinghua University, Beijing 100084, China
[3] College of Physics, Sichuan University, Chengdu 610065, Sichuan, China

**Abstract** The analyses of the LHCb data on $X(2900)$ in the $D^-K^+$ spectrum are performed. Both dynamically generated and explicitly introduced $X_1(2900)$ are considered. The results show that both these two approaches support the interpretation of $X_1(2900)$ as a $\bar{D}_1 K$ molecular state, with $J^P = 1^-$ and an iso-singlet interpretation is much more favorable. The effect of triangle singularity on the production of $X_1(2900)$ is also discussed, and it is found that it cannot be interpreted as a pure triangle cusp.

## 1 Introduction

In year 2020, LHCb collaboration declared that two resonant structures, called $X_0(2900)$ and $X_1(2900)$, were found in the $D^-K^+$ invariant mass spectrum via $B^\pm \to D^+D^-K^\pm$. The masses and widths of two resonances are [1],

$m_{X_0(2900)} = 2.866 \pm 0.007 \pm 0.002$ GeV,

$\Gamma_{X_0(2900)} = 57 \pm 12 \pm 4$ MeV,

and

$m_{X_1(2900)} = 2.904 \pm 0.005 \pm 0.001$ GeV,

$\Gamma_{X_1(2900)} = 110 \pm 11 \pm 4$ MeV,

respectively. The two resonances can both decay into the $D^-K^+$ final state, so their valence quarks are $ud\bar{s}\bar{c}$, i.e., the valence quarks have different flavors. These structures arise many research interests.

Recently, there are many investigations on the nature of $X_1(2900)$ and $X_0(2900)$. For example, SU(3) flavor symmetry being considered in Ref. [2], the newly observed structures are contained in the irreducible representation $\bar{6}$ and 15. Then, diquark model with two-body Columb interactions and chromomagnetic interactions are introduced. Mass spectrum and decay properties of open-charm tetraquark are calculated within this scheme. In the end, the results suggest that $X_0(2900)$, $X_1(2900)$ can be seen as radial excited tetraquark with $J^{PC} = 0^+$ and orbitally excited tetraquark with $J^P = 1^-$, respectively. Of course, there also exist analyses where $X_{0,1}(2900)$ are interpreted as a triangle cusp. For example, the process $B^+ \to D^-D^+K^-$ is constructed by a nonrelativistic triangle diagram and combined with final state interactions in Ref. [3]. As a conclusion, $X_1(2900)$ is thought to be a $\bar{D}^*K^*$ or $\bar{D}_1 K$ triangle cusp, but within the final state interaction, the molecular state with valence quarks $ud\bar{s}\bar{c}$ could also exist. In Ref. [4], the production of $X(2900)$ by means of triangle diagrams is also investigated. The analyses show that $X_0(2900)$, $X_1(2900)$ can be produced via the $\chi_{c1} K^{*+} D^{*-}$ and the $D_{sJ}^{*+} \bar{D}_1^0 K^0$ triangle loops, respectively.

[a] e-mail: haochen0393@pku.edu.cn
[b] e-mail: qihongrong@tsinghua.edu.cn (corresponding author)
[c] e-mail: zhenghq@scu.edu.cn

Except for these interpretations, more researchers prefer to consider $X_{0,1}(2900)$ as hadronic molecules. For instance, based on one-boson exchange model [5], $X_0(2900)$ is considered as a $D^*\bar{K}^*$ molecular state with $IJ^P = 00^+$, while $X_1(2900)$ is not supposed to be a molecular state. With sloving Lippmann-Schwinger equation and quasipoteintial Bethe-Salpeter equation (qBSE) [6–8], $X_0(2900)$ and $X_1(2900)$ are interpreted as $\bar{D}^*K^*$ and $\bar{D}_1 K$ molecular states, respectively. In Quark delocalization color screening model (QDCSM) [9], energy spectra of tetraquarks composed of $ud\bar{s}\bar{c}$ with meson-meson and diquark-antidiquark structures show that $X_0(2900)$ could be the candidate of $\bar{D}^*K^*$ molecular state, $IJ^P = 00^+$. Moreover, the calculation by means of QCD sum rule suggests $X_0(2900)$ to be a $\bar{D}^*K^*$ molecular state [10,11], and $X_1(2900)$ to be a $P$-wave compact tetraquark state [10].

Although many investigations have been attempted to explain the nature of $X_{0,1}(2900)$, only a few of them made fits to the experimental data. In order to gain a deeper insight into the nature of $X_{0,1}(2900)$, it is meaningful to do further researches on the newly observed spike, $X_{0,1}(2900)$, by LHCb collaboration. By the mean time, $X_1(2900)$ is mea-sured to be the predominant signal, but $X_0(2900)$ have poor significance [1]. Hence only $X_1(2900)$ is performed in the fits of this analysis. Then, two approaches are introduced. Firstly, a dynamically generated $X_1(2900)$ is under consideration, and couple-channel K-matrix approach for $\bar{D}K$, $\bar{D}_1K$ scattering is introduced. Secondly, a Flatté-like formula is employed to describe an explicit $X_1(2900)$. After getting the fit parameters, pole positions of the amplitudes are searched for in the complex $s$ plane. Pole counting rule (PCR) [12] is used to discriminate whether $X_1(2900)$ is a hadron molecule or a compact tetraquark state. In the end, a discussion on the contribution of triangle cusp to the $X_1(2900)$ production is also performed.

Finally, the results suggest that $X_1(2900)$ is a $\bar{D}_1K$ molecule and the fits using triangle diagram can not reproduce a correct $X_1(2900)$ peak, i.e., $X_1(2900)$ is not like a triangle cusp.

## 2 Dynamically generated $X_1(2900)$

In this section, $\bar{D}K$, $\bar{D}_1K$ perturbative scattering amplitudes will be constructed using heavy meson chiral perturbation theory (HM$\chi$PT) [13]. Then, couple-channel K-matrix approach is employed to get the unitarized amplitudes. In order to identify the quantum numbers of $X_1(2900)$, it is assumed that the resonance is produced by the final state interactions (FSI) with special quantum numbers and the contributions from FSI with other quantum numbers are smooth. Poles of unitarized amplitudes will be searched for in the complex $s$ plane and PCR, which is widely used in many studies [14–17] will also be employed in this work to investigate the nature of $X_1(2900)$.

### 2.1 Perturbative amplitudes

Interactions between mesons composed of a heavy quark ($c$, $b$) and a light quark ($u$, $d$, $s$) and light pesudoscalar or vector mesons can be described by means of HM$\chi$PT. When the heavy quark mass tends to be infinite and the light quark tends to be massless, the spin-flavor symmetry and chiral symmetry will occur in QCD. Then, the HM$\chi$PT is constructed from these conditions.

In the HM$\chi$PT, heavy meson fields are represented by covariant, $4 \times 4$ Dirac-type matrices. For example, the heavy meson fields composed of a heavy anti-quark $\bar{Q}$ and a light quark $q$ with $J^P = 0^-$, $1^-$ can be defined as [18]:

$$H_a^{(\bar{Q})} = \left[ P_{a\mu}^{*(\bar{Q})}\gamma^\mu - P_a^{(\bar{Q})}\gamma_5 \right] \frac{1-\slashed{v}}{2}. \tag{1}$$

It is a linear combination of a pseudoscalar field $P_a^{(\bar{Q})}$ and a vector field $P_{a\mu}^{*(\bar{Q})}$, where $a$ is the flavor of light quarks, $a = 1, 2, 3$ for $u$, $d$, $s$ quark. $P_{a\mu}^{*(\bar{Q})}$, $P_a^{(\bar{Q})}$ annihilate the heavy meson states composed of $\bar{Q}q_a$, $J^P = 1^-$, $0^-$, respectively. In this paper, $\bar{Q} \to \bar{c}$, so $P_{a\mu}^{*(\bar{Q})}$, $P_a^{(\bar{Q})}$ are to annihilate $\bar{D}^*$, $\bar{D}$, respectively. The 4-velocity $v$ of the heavy meson, is conserved in strong interaction processes and it satisfies $v^\mu P_{a\mu}^* = 0$, $v \cdot v = 1$. It should be noted that every heavy meson field $P$ contains a normalized factor $\sqrt{M}$ [1], where $M$ is mass of the heavy meson.

Then, the conjugate field, which creates heavy-light mesons containing a heavy quark $\bar{Q}$ and a light quark $q$, is defined as,

$$\bar{H}_a^{(\bar{Q})} = \gamma_0 H_a^{(\bar{Q})\dagger} \gamma_0 = \frac{1-\slashed{v}}{2} \left[ P_a^{*(\bar{Q})\mu\dagger}\gamma_\mu + P_a^{(\bar{Q})\dagger}\gamma_5 \right]. \tag{2}$$

In the same way, the field, which annihilates heavy-light mesons composed of $\bar{Q}q$ with $J^P = 1^+$, $2^+$ is defined as,

$$T_a^{(\bar{Q})\mu} = \left[ P_{2a}^{(\bar{Q})\mu\nu}\gamma_\nu - \sqrt{\frac{3}{2}} P_{1a\nu}^{(\bar{Q})} \gamma_5 \left( g^{\mu\nu} - \frac{1}{3}(\gamma^\mu - v^\mu)\gamma^\nu \right) \right] \times \frac{1-\slashed{v}}{2}. \tag{3}$$

and its conjugate field is defined as,

$$\bar{T}_{a\mu}^{(\bar{Q})} = \gamma_0 T_{a\mu}^{(\bar{Q})\dagger} \gamma_0 = \frac{1-\slashed{v}}{2} \times \left[ P_{2a\mu\nu}^{(\bar{Q})\dagger}\gamma^\nu + \sqrt{\frac{3}{2}} P_{1a}^{(\bar{Q})\nu\dagger} \left( g_{\mu\nu} - \frac{1}{3}\gamma_\nu(\gamma_\mu - v_\mu) \right) \gamma_5 \right]. \tag{4}$$

If $\bar{Q} \to \bar{c}$, $P_{1a}^{(\bar{Q})\nu}$, $P_{2a}^{(\bar{Q})\mu\nu}$ ($P_{1a}^{(\bar{Q})\nu\dagger}$, $P_{2a}^{(\bar{Q})\mu\nu\dagger}$) annihilate (create) $\bar{D}_1$, $\bar{D}_2^*$, respectively.

The interaction lagrangians between the heavy meson and pseudoscalar meson octet can be constructed as, e.g. see Ref. [18],

$$\begin{aligned}
\mathcal{L}_{\bar{D}\bar{D}\phi\phi} &= -i\beta \left\langle \bar{H}_a^{(\bar{Q})} v^\mu (\mathcal{V}_\mu)_{ab} H_b^{(\bar{Q})} \right\rangle, \\
\mathcal{L}_{\bar{D}_1\bar{D}_1\phi\phi} &= -i\beta_2 \left\langle \bar{T}_{a\lambda}^{(\bar{Q})} v^\mu (\mathcal{V}_\mu)_{ab} T_b^{(\bar{Q})\lambda} \right\rangle, \\
\mathcal{L}_{\bar{D}\bar{D}_1\phi\phi} &= -i\zeta_1 \left\langle \bar{H}_a^{(\bar{Q})} (\mathcal{V}_\mu)_{ab} T_b^{(\bar{Q})\mu} \right\rangle + h.c.,
\end{aligned} \tag{5}$$

from which the contact interactions between heavy mesons and light pesudosaclar mesons can be deduced. In Eq. (5), $\langle \cdots \rangle$ means trace over the gamma matrices. $\mathcal{V}_\mu = \frac{1}{2}(\xi^\dagger \partial_\mu \xi + \xi \partial_\mu \xi^\dagger)$, $\xi = \exp(i\phi/f_\pi)$ and $f_\pi = 132$ MeV is the pion decay constant. $v_\mu$ should be replaced by $i \overset{\leftrightarrow}{\partial_\mu}$ $/\sqrt{M_a M_b}$, which means the partial derivate of heavy meson field and $\overleftrightarrow{\partial}_\mu = \overrightarrow{\partial}_\mu - \overleftarrow{\partial}_\mu$. $\phi$ is the matrix of pseudoscalar mesons,

$$\phi = \begin{pmatrix} \frac{1}{\sqrt{2}}\pi^0 + \frac{1}{\sqrt{6}}\eta & \pi^+ & K^+ \\ \pi^- & -\frac{1}{\sqrt{2}}\pi^0 + \frac{1}{\sqrt{6}}\eta & K^0 \\ K^- & \bar{K}^0 & -\frac{2}{\sqrt{6}}\eta \end{pmatrix}. \quad (6)$$

---

[1] In the conventional quantum field theory, the normalization of state vector is $\langle H(p')|H(p)\rangle = 2E_p(2\pi)^3\delta^3(\mathbf{p} - \mathbf{p}')$, while in HM$\chi$PT, the normalization of heavy meson is $\langle H(v', k')|H(v, k)\rangle = 2v^0\delta_{v,v'}(2\pi)^3\delta^3(\mathbf{k} - \mathbf{k}')$. The relationship between the two formulae is $|H(p)\rangle = \sqrt{m_H}[|H(v)\rangle + O(1/m_Q)]$. One is referred to [13] for more details.

Considering the isospin symmetry, one has,

$$\begin{aligned}
|K\bar{D}, I=0\rangle &= \frac{1}{\sqrt{2}}\left(|K^+ D^-\rangle - |K^0 \bar{D}^0\rangle\right), \\
|K\bar{D}, I=1\rangle &= \frac{1}{\sqrt{2}}\left(|K^+ D^-\rangle + |K^0 \bar{D}^0\rangle\right), \\
|K\bar{D}_1, I=0\rangle &= \frac{1}{\sqrt{2}}\left(|K^+ D_1^-\rangle - |K^0 \bar{D}_1^0\rangle\right), \\
|K\bar{D}_1, I=1\rangle &= \frac{1}{\sqrt{2}}\left(|K^+ D_1^-\rangle + |K^0 \bar{D}_1^0\rangle\right).
\end{aligned} \quad (7)$$

So, the perturbative amplitudes with given isospin can be deduced,

$$\begin{aligned}
i\mathcal{M}^I_{\bar{D}K \to \bar{D}K} &= \theta_I \frac{i\beta}{f_\pi^2}(s-u), \\
i\mathcal{M}^I_{\bar{D}_1 K \to \bar{D}_1 K} &= \theta_I \frac{i\beta_2}{f_\pi^2}\epsilon^*(p_3) \cdot \epsilon(p_1)(s-u). \\
i\mathcal{M}^I_{\bar{D}K \to \bar{D}_1 K} &= \theta_I i\sqrt{\frac{2}{3}}\zeta_1 \frac{\sqrt{m_D m_{D_1}}}{f_\pi^2}\epsilon^*(p_3) \cdot (p_2+p_4),
\end{aligned} \quad (8)$$

where $p_1$, $p_3$ are momentum of heavy mesons and $p_2$, $p_4$ are momentum of $K$. $\theta_I$ is a isospin factor, $\theta_I = 1$ for $I=0$ and $\theta_I = -1$ for $I=1$. The $s$, $u$ in Eq. (8) are Mandelstam variables, where $s = (p_1+p_2)^2$, $u = (p_1-p_4)^2$. It can be seen that Weinberg–Tommozawa term occurs when only contact interactions are considered.

2.2 Partial wave perturbative amplitudes

Then, helicity amplitudes are calculated. In order to analyze the quantum number of the resonance, one should do the partial wave expansion for helicity amplitudes [19],

$$\begin{aligned}
&\left\langle \lambda_3 \lambda_4 \left| \mathcal{M}^{IJ}(E_p) \right| \lambda_1 \lambda_2 \right\rangle \\
&= \frac{1}{2\pi}\int_{-1}^{1}\left\langle \Omega' \lambda_3 \lambda_4 | \mathcal{M}^I(E_p, \mathbf{p}, \mathbf{p}')|\Omega \lambda_1 \lambda_2 \right\rangle d^J_{\lambda,\lambda'}(z_s) dz_s,
\end{aligned} \quad (9)$$

where $z_s = \cos\theta$, where $\theta$ is the scattering angle. $d^J_{\lambda,\lambda'}$ is the Wigner d-function. $\lambda = \lambda_1 - \lambda_2$, $\lambda' = \lambda_3 - \lambda_4$, and $\lambda_{1,2}(\lambda_{3,4})$ are helicities of incoming (outgoing) particles. The

Mandelstam variables $t$, $u$ can be represented by $s$ and $z_s$,

$$\begin{aligned}
t =& m_1^2 + m_3^2 - \frac{(s+m_1^2-m_2^2)(s+m_3^2-m_4^2)}{2s} \\
&+ \sqrt{[s-s_{L,12}][s-s_{R,12}][s-s_{L,34}][s-s_{R,34}]}\frac{z_s}{2s}, \\
u =& m_1^2 + m_4^2 - \frac{(s+m_1^2-m_2^2)(s+m_4^2-m_3^2)}{2s} \\
&- \sqrt{[s-s_{L,12}][s-s_{R,12}][s-s_{L,34}][s-s_{R,34}]}\frac{z_s}{2s},
\end{aligned} \quad (10)$$

where $s_{L,ij} = (m_i - m_j)^2$, $s_{R,ij} = (m_i + m_j)^2$.

After doing the integral in Eq. (9), the helicity amplitudes with given angular momentum $J$ can be work out. But in practice, it is convenient to discuss resonances' properties in angular momentum bases than helicity basis, so partial wave amplitudes in Eq. (9) should be transformed into angular momentum basis. The details of the basis transformation and the final expressions of amplitudes in angular momentum basis are listed in Appendix A.

2.3 Unitarized amplitudes

The couple-channel K-matrix method will be introduced to unitarize the partial-wave perturbative amplitudes. From the unitary condition,

$$\text{Im } T^{-1}(s) = -\rho(s), \quad (11)$$

the imaginary part of the inversed amplitude is determined. If the real part of inversed amplitude is noted as $\mathcal{K}^{-1}$ and combining $\mathcal{K}^{-1}$ with Eq. (11), it gives,

$$T^{-1} = -i\rho(s) + \mathcal{K}^{-1}. \quad (12)$$

Here $T$ is a matrix with dimension n, where n is the number of channels. $\mathcal{K}$ is a real symmetric matrix above the unitary threshold and $\rho = \text{diag}\{\rho_1, \ldots, \rho_n\}$, is the diagnoal matrix of phase space factors. Then, it can be deduced from Eq. (12),

$$T = \mathcal{K} \cdot [1 - i\rho(s)\mathcal{K}]^{-1}. \quad (13)$$

$\mathcal{K}$ is a real symmetric matrix. In this work, channels $\bar{D}K$ and $\bar{D}_1 K$ are under consideration, so $\mathcal{K}$ can be written as,

$$\begin{pmatrix} \mathcal{T}_{\bar{D}K \to \bar{D}K} & \mathcal{T}_{\bar{D}K \to \bar{D}_1 K} + \mathcal{P}_{12} \\ \mathcal{T}_{\bar{D}_1 K \to \bar{D}K} + \mathcal{P}_{12} & \mathcal{T}_{\bar{D}_1 K \to \bar{D}_1 K} \end{pmatrix}^{IJ} \\
\equiv \begin{pmatrix} \mathcal{K}_{11} & \mathcal{K}_{12} \\ \mathcal{K}_{21} & \mathcal{K}_{22} \end{pmatrix}^{IJ}, \quad (14)$$

where $\mathcal{T}$s are amplitudes in angular momentum basis, one is referred to Appendix A for more details. $\mathcal{K}_{12} = \mathcal{K}_{21}$ and $\mathcal{P}_{12} = c_{012} + c_{112}(\sqrt{s} - m_{\text{th}2})$ with $m_{\text{th}2} = (m_{D_1} + m_K)$, is a first order polynomial of c.m. energy. The polynomial is to simulate the effects from higher order interactions such as vector meson exchanging diagrams on physical region. In principle, polynomials can be added to every element of $\mathcal{K}$ matrix, but in practice, the results show that adding polynomials only to non-diagnoal elements can fit good enough.

The full amplitude of process $B \to D\bar{D}K$ can be represented as,

$$\mathcal{A} = \alpha_1(s)T_{11}(s) + \alpha_2(s)T_{21}(s) \quad (15)$$

where $\alpha_{1,2}$ are the polynomials of $s$. In practice, the forms of $\alpha_{1,2}$ are taken as, $\alpha_1 = a_{10} + a_{11}(\sqrt{s} - (m_{\bar{D}} + m_K))$, $\alpha_2 = a_{20}$. The amplitude in Eq. (15) satisfies the theorem of FSI [20],

$$\text{Im}\,\mathcal{A}_k(s) = \sum_l \mathcal{A}_l^*(s)\rho_l(s)T_{lk}(s), \quad (16)$$

where the $l$, $k$ represent scattering or decay channels.

It is assumed that the resonance is only produced by the FSI of $\bar{D}K$, $\bar{D}_1 K$ with $J^P = 1^-$. The contributions from the FSI with other quantum numbers are smooth and can be absorbed into the background.

Hence, the invariant mass spectrum can be parameterized as [21,22],

$$\frac{d\sigma}{d\sqrt{s}} = \frac{1}{(2\pi)^3}\frac{1}{32m_B^3}\int 2\sqrt{s}|\mathcal{A}(s,s')|^2 ds' + b_0\rho(s)_{\bar{D}K}. \quad (17)$$

where $s' = (p_D + p_K)^2$, $s = (p_{\bar{D}} + p_K)^2$ and $p_K$, $p_D$, $p_{\bar{D}}$ are momentum of outgoing particles $K$, $D$, $\bar{D}$ written in the c.m. frame of $\bar{D}K$ scattering. $b_0\rho(s)_{\bar{D}K}$ represents the incoherent background contributions.

## 2.4 Fit results and discussion

The value of parameter $\beta$ is estimated to be 0.9 using vector meson dominance model [23]. In fits, it is fixed to 0.9 as in Refs. [18,23] and other parameters are regarded as free parameters since there are no accepted estimates about them. The fit curves are listed in Fig. 1 and fit parameters are listed in Table 1.

From Fig. 1, it can be seen that both the FSI with $IJ^P = 01^-$ and $IJ^P = 11^-$ can produce $X_1(2900)$. The poles near the threshold of $\bar{D}_1 K$ in complex $s$ plane are searched for. Different Riemann sheets are identified by signs of phase space factors and the definition is listed in Table 2. The fit results show that the FSI with $IJ^P = 01^-$ and $IJ^P = 11^-$ both generate only one pole near the threshold of $\bar{D}_1 K$ in the sheet II. The pole positions are $(2.928 - 0.034i)$ GeV for $IJ^P = 01^-$ and $(2.925 - 0.039i)$ GeV for $IJ^P = 11^-$. The distant poles which are about 3~5 times of the lineshape width of the resonance away from the threshold may be found, but they are irrelevant since they are too far away. According to PCR [12], $X_1(2900)$ should be recognized as a molecular state if there is only one pole near the threshold. So, the method used in this section dynamically generates a pole near $\bar{D}_1 K$ threshold to fit the data, which means that $X_1(2900)$ is a hadronic molecule of $\bar{D}_1 K$, $J^P = 1^-$.

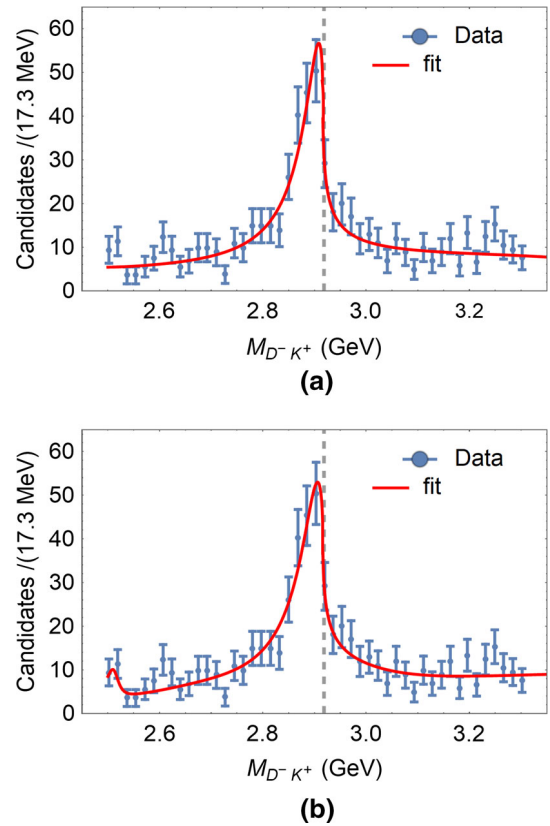

**Fig. 1** The fit results using amplitude with final state interactions whose quantum numbers are $IJ^P = 01^-$ (a) and $IJ^P = 11^-$ (b). The fit goodness are $\chi^2/d.o.f. = 42.95/40$ and $\chi^2/d.o.f. = 37.94/39$, respectively. The dashed vertical line is located at $\bar{D}_1 K$ threshold. Data are from Ref. [1]

**Table 1** Parameters of two fits with dynamically generated $X_1(2900)$ where $a'_{20} = a_{20}/a_{10}$, $a'_{11} = a_{11}/a_{10}$

| Parameters | $IJ^P = 01^-$ Values | $IJ^P = 11^-$ |
|---|---|---|
| $\beta_2$ | $-0.21 \pm 0.03$ | $-0.26 \pm 0.07$ |
| $\zeta_1$ | $-0.90 \pm 0.48$ | $5.85 \pm 2.00$ |
| $c_{012}$ | $25.25 \pm 2.48$ | $70.00 \pm 27.87$ |
| $c_{112}$ | $35.04 \pm 8.46$ | $71.37 \pm 35.89$ |
| $a_{10}$ | $35.42 \pm 5.58$ | $14.41 \pm 3.00$ |
| $a'_{20}$ | $-6.30 \pm 1.72$ | $10.00 \pm 2.13$ |
| $a'_{11}$ | fixed=0 | $4.09 \pm 1.63$ |
| $b_0$ | $10.15 \pm 2.92$ | $13.70 \pm 1.29$ |

A short discussion about values of parameter $\zeta_1$ can be performed. In principle, it should be fixed from experimental data, e.g. the decay of $D_1 \to D\rho(\omega)$. But such a process is forbidden kinetically [24]. In Ref. [24], it is estimated that $\zeta_1 \approx \pm 0.16$ from the decay of $K_1 \to K\rho$. But in Table 1, it

| Table 2 Definition of Riemann sheets | I | II | III | IV |
|---|---|---|---|---|
| $\rho_1$ | + | − | − | + |
| $\rho_2$ | + | + | − | − |

can be seen that the value of $\zeta_1$ in the fit with $IJ^P = 11^-$ is much larger than the estimated value in Ref. [24]. In fact, if $\zeta_1$ is restricted to be small, e.g. in $[-1, 1]$ during the fit where the $IJ^P$ of FSI is $11^-$, there will almost inevitably exist poles near the threshold in the first Riemann sheet and this is unacceptable. It suggests that $X_1(2900)$ is more leaning toward an iso-singlet than an iso-triplet using current data. Furthermore, as in Ref. [25], we also suggest experimental groups to measure processes such as $B^+ \to D^0 X^+$, $X^+ \to \bar{D}^0 K^+$ or $B^0 \to D^+ X^-$, $X^- \to D^- K^0$, which are the possible productive channels of isospin partners of $X_1(2900)$ to determine its isospin.

## 3 Explicitly introduced $X_1(2900)$

In this section, the explicitly introduced $X_1(2900)$ is considered. Using Flatté-like parametrization, the analysis of the experimental data is performed. After doing these, poles of the amplitude will be searched for in the complex $s$ plane.

The amplitude in Flatté-like formula can be written as,

$$\mathcal{M} = \frac{g \cdot n_{\bar{D}K}(s)}{s - M_X^2 + iM_X \left[ g_1 \rho_{\bar{D}K}(s) n_{\bar{D}K}^2(s) + g_2 \rho_{\bar{D}_1 K}(s) \right]}, \quad (18)$$

with,

$$n(s) = \left( \frac{p}{p_0} \right)^l F_l (p/p_0), \quad (19)$$

where $l$ is the quantum number of orbital angular momentum in the two-body channel and $p$ is the center-of-mass momentum of one daughter particle in this channel [^2]. $p_0$ is a momentum scale and $F_l$ is a form factor which can be written as $F_0(z) = 1$, $F_1(z) = \sqrt{1/(1+z)}$, $F_2(z) = \sqrt{1/(9 + 3z + z^2)}$ for $l = 0, 1, 2$, respectively [26] with $z = (p/p_0)^2$. $\rho(s)$ is the two-body phase space of the final state and $M_X$ is the bare mass of $X_1(2900)$.

As shown in Eq. (18), the situation that $X_1(2900)$ couples to $S$-wave $\bar{D}_1 K$ and $P$-wave $\bar{D} K$ channels with $J^P = 1^-$ is investigated. Then, the $D^- K^+$ invariant mass spectrum can be parameterized as,

$$\frac{d\sigma}{d\sqrt{s}} = p_{\bar{D}K} \cdot |\mathcal{M}(s)|^2 + b_0 \rho_{\bar{D}K}(s)., \quad (20)$$

[^2]: For a two-body final state with $m_a$, $m_b$, $p = \sqrt{[s-(m_a+m_b)^2][s-(m_a-m_b)^2]}/2\sqrt{s}$. If $p^2 < 0$, it can be analytic continued as $p = i\sqrt{-p^2}$.

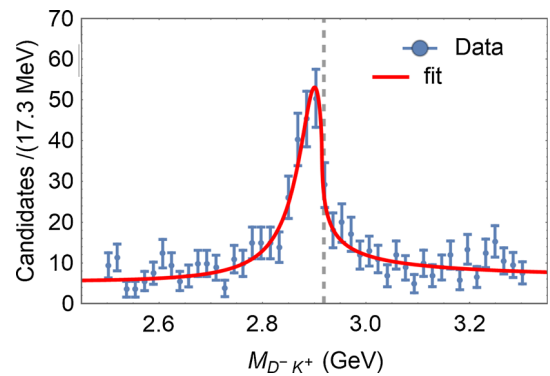

Fig. 2 Fit result with explicit $X_1(2900)$. The dashed vertical line is located at the threshold of $\bar{D}_1 K$. Data are from Ref. [1]

Table 3 Parameters of fit with explicit $X_1(2900)$

| Parameters | Value | Standard error |
|---|---|---|
| $M_X$ (GeV) | 2.837 | 0.015 |
| $g_1$ (GeV) | 1.550 | 0.764 |
| $g_2$ (GeV) | 1.850 | 1.055 |
| $p_0$ (GeV) | 1.194 | 0.119 |
| $g$ | 9.632 | 0.645 |
| $b_0$ | 11.072 | 0.958 |

Table 4 Pole positions of the fit with explicit $X_1(2900)$. The pole located in sheet III is too far from the threshold of $\bar{D}_1 K$, so there is only one pole near the threshold

| Riemann sheet | I | II | III | IV |
|---|---|---|---|---|
| Pole position (GeV) | – | $2.910 - 0.039i$ | $2.435 - 0.012i$ | – |

where the last term above represents the incoherent background.

The fit result shows that the experimental data can also be fit well using Flatté-like parametrization. The fit result is in Fig. 2 and the parameters as well as their values are listed in Table 3.

Then, the poles of the amplitude are searched for in complex $s$ plane. The definition of Riemann sheets is listed in Table 2 and the positions of poles is listed in Table 4.

From Table 4, it is found that only the pole located in sheet II is near the threshold of $\bar{D}_1 K$. The pole, located in sheet III, is about 5 times of the lineshape width away from the threshold and should not be recognized as a pole near the threshold [16]. According to PCR, it is to be recognized as a molecular state if there is only one pole near the threshold. Then a conclusion can be drawn that $X_1(2900)$ is a molecular state of $\bar{D}_1 K$ with $J^P = 1^-$ in the scheme of explicitly introduced resonance.

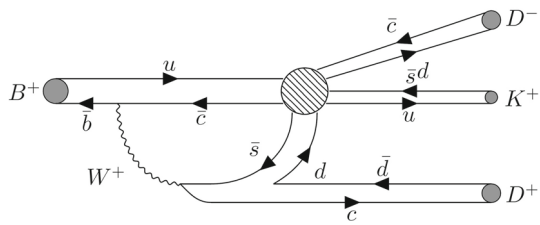

**Fig. 3** Schematic diagram of process $B^+ \to D^+D^-K^+$. The shadow circle represents the production vertex of $X_1(2900)$

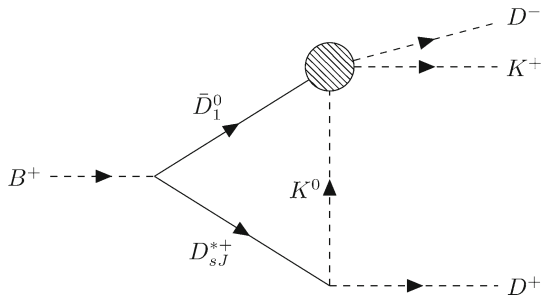

**Fig. 4** Triangle diagram of the process $B^+ \to D^+D^-K^+$. The shadow circle represents the production vertex of $X_1(2900)$

## 4 Some remarks on production through triangle cusp

As mentioned in Introduction, some studies supposed that $X_1(2900)$ is produced through a triangle cusp [3,4]. In this section, the possibility of $X_1(2900)$ to be a triangle cusp also be simply discussed for completeness. Firstly, it is supposed that $B$ meson decay into a charmed strange meson and a anti-charmed meson through Cabibbo-favored process. Then the final state $D^+D^-K^+$ can be obtained by these two intermediate particles' rescattering. Figure 3 shows a $B$ decay mode through Cabibbo-favored process.

In this section, such a triangle diagram to produce the final state $D^+D^-K^+$: $B \to \bar{D}_1 + D^*_{sJ}$, $D^*_{sJ} \to K + D$, $\bar{D}_1 + K \to \bar{D}K$ is considered, as shown in Fig. 4.

Interaction lagrangians involved are listed as follows,

$$\begin{aligned}
\mathcal{L}_{BD^*D} &= g_B B^+ D^{*-}_{sJ\mu} D^{0\mu}_1, \\
\mathcal{L}_{D^*DK} &= g_1 D^{*+\mu}_{sJ} \partial_\mu D^- \bar{K}^0 + g_2 D^{*+\mu}_{sJ} D^- \partial_\mu \bar{K}^0, \\
\mathcal{L}_{DD_1K} &= g \bar{D}^{0\mu}_1 D^+ [K^0 \partial_\mu K^- - \partial_\mu K^0 K^-] \\
&\quad + g' \bar{D}^{0\mu}_1 \partial_\mu D^+ K^0 K^-.
\end{aligned} \tag{21}$$

The momentum of $B^+$, $D^-$, $K^+$, $D^+$ are donated by $P$, $q_1$, $q_2$, $q_3$, respectively. Let $q_{ij} = (q_i + q_j), i, j = 1, 2, 3$. Usually, the propagator of vector particle is written in unitary gauge in phenomenological studies. But in practice, it is found that the spike will not be obvious at all if the propagator of vector is written in unitary gauge. In view of this, the propagator of vector meson is written in Landau

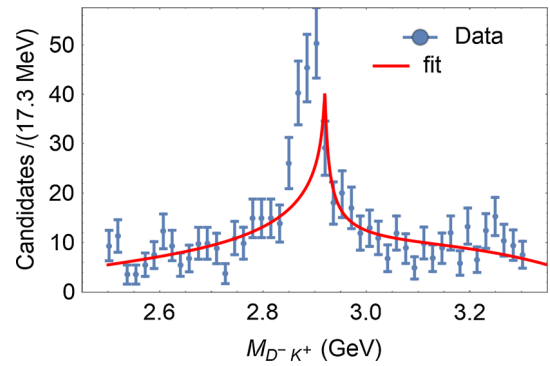

**Fig. 5** Fit result of a single triangle diagram. Data are from Ref. [1]

gauge,

$$\frac{-ig_{\mu\nu}}{k^2 - m^2 + i\varepsilon}. \tag{22}$$

Then the amplitude of triangle diagram can be written as,

$$\begin{aligned}
i\mathcal{M}_{\text{triangle}} &= \int \frac{d^D l}{(2\pi)^D} \frac{(g_B g^{\mu\nu})(g_1 q_3 + g_2 l)_\sigma g^{\nu\sigma}[g(q_2 + l)_\rho + g' q_{1\rho}]g^{\mu\rho}}{[(l + q_3)^2 - m^{*2} + i\varepsilon][l^2 - m^2_K + i\varepsilon]} \\
&\quad \times \frac{1}{[(l - q_{12})^2 - m^2_{D_1} + i\varepsilon]}.
\end{aligned} \tag{23}$$

where $m^*$ is the mass of $D^*_{sJ}$, which is argued to be $D^*_{s1}(2860)$ or $D^*_{s1}(2700)$ in Ref. [4]. In the fit below, $m^*$ is regarded as a free parameter. Then the amplitude in Eq. (23) is evaluated [27–29] and the three-point loop integrals involved are listed in Appendix B.

After taking incoming and outgoing particles on their mass shell, i.e. let $P^2 = m^2_B, q^2_1 = q^2_3 = m^2_D, q^2_2 = m^2_K$, one can find that the amplitude is a function of $q_{12}$ and $q_{23}$. Because $X_1(2900)$ was observed in $D^-K^+$ final state, so $q_{23}$ should be integrated out [21]. At last, the invariant mass spectrum is parameterized as,

$$\begin{aligned}
\frac{d\sigma}{dq_{12}} &= \frac{1}{(2\pi)^3} \frac{1}{32m^3_B} \\
&\quad \times \int 2q_{12} |\mathcal{M}_{\text{triangle}}(q_{12}, q_{23})|^2 dq^2_{23} + b_0 \rho_{\bar{D}K}(s),
\end{aligned} \tag{24}$$

In the fit, a finite width is given to $m^*$ to simulate physical situation of $D^*_{s1}(2860)$ or $D^*_{s1}(2700)$, i.e. taking place of $m^*$ by $m^* - i\Gamma$, where $\Gamma$ is the half-width of $D^*_{sJ}$. As for $\bar{D}^0_1$, because of its small width (about 30 MeV), so its width can be ignored in the fit. The fit result is shown in Fig. 5.

It gives the parameter $m^* = 2.544 \pm 0.097$ GeV, $\Gamma = 0.048 \pm 0.059$ GeV. The mass parameter is much smaller than $D^*_{s1}(2860)$ but kind of like $D^*_{s1}(2700)$ within error bars. The fit curve in Fig. 5 shows that a pure triangle diagram can indeed produce a cusp near the threshold, but it can not explain such a broad structure observed by experiment. Of course, some form factor may also be added to the vertex of $X_1(2900)$ production, e.g. [3]

$$F(x) = \frac{1}{1+x^2}, \quad x = \frac{p}{\Lambda},$$

where $\Lambda$ is a momentum scale. But the result shows it can not improve fit goodness.

A detailed discussion about the triangle singularity induced by the three-point loop integral is performed in Appendix B. As the result, the particle $D_{sJ}^*$ is neither likely to be $D_{s1}^*(2700)$ nor $D_{s1}^*(2860)$ which is the candidate of $D_{sJ}^*$ in Ref. [4]. From the numerical results, if a spike was produced near 2.90 GeV, $m^*$ should take the value of about 2.5 GeV which is far away from the mass of $D_{s1}^*(2700)$ or $D_{s1}^*(2860)$. Nevertheless, it still can not fit the data if $m^*$ took the value about 2.5 GeV which is shown in Fig. 5.

So, it is suggested that $X_1(2900)$ can hardly be produced by a triangle cusp. In other words, at least, the effect of triangle cusp looks not to be the leading contribution in $X_1(2900)$ production.

## 5 Conclusion

In this paper, the molecular picture of $X_1(2900)$ is investigated. Both the data-based analyses of dynamically generated hadron scattering state by couple-channel K-matrix approaches, and an explicitly introduced resonance by Flatté-like parameterization support this point of view. $X_1(2900)$ is leaning toward an iso-singlet than an iso-triplet by comparing the value of fit parameter $\zeta_1$ with the estimated value in Ref. [24]. The conclusion in this work that $X_1(2900)$ is a $\bar{D}_1 K$ molecule with $J^P = 1^-$ confirms the points of view in some earlier studies [6–8]. In the end of this work, the possibility of $X_1(2900)$ to be a triangle cusp is also considered. Fitting with a triangle diagram shows that it can not produce such a broad structure and direct analyses of three-point loop integral singularities implies a correct $X_1(2900)$ can not be reproduced within reasonable range of parameters. This confirms the conclusion of $X_1(2900)$ to be a hadronic molecule again.

It is noticed that in Ref. [30], $X_0(2900)$ and $X_1(2900)$ are interpreted as $S$-wave and $P$-wave $\bar{D}^* K^*$ molecules, respectively. Although data can be fit well in the Ref. [30], it should be noteworthy that almost all heavy exotic hadrons that have been observed so far locate near the threshold of two hadrons and can be attributed to the $S$-wave hadronic molecules (except for $Z_c(4430)$ which is recognized as a $P$-wave molecule [31]). To a large extend, an $S$-wave $\bar{D}_1 K$ molecule picture in this paper is more acceptable for $X_1(2900)$.

Duing to lack of data, the existence of $X_0(2900)$ is less significant so it's hard to investigate $X_0(2900)$ using data-based methods. Hence, we also suggest experiments to measure processes, e.g. $B^+ \rightarrow D^0 X^+, X^+ \rightarrow \bar{D}^0 K^+$ or $B^0 \rightarrow D^+ X^-, X^- \rightarrow D^- K^0$ as suggested in Ref. [25] to determine the isospin of $X_1(2900)$ and the existence of $X_0(2900)$.

**Acknowledgements** This work is supported in part by National Nature Science Foundations of China under Contract Number 11975028 and 10925522; and China Postdoctoral Science Foundation under Contract Number 2020M680500.

**Data Availability Statement** This manuscript has no associated data or the data will not be deposited. [Authors' comment: There is no data to be deposited since all of data in our article are measured by LHCb [Phys. Rev. D 102, 605 112003 (2020)].]

## Appendix A: Parital wave amplitudes in angular momentum bases

The relations between amplitudes in helicity basis and angular momentum basis are given by [19,32],

$$\begin{aligned}
\mathcal{T}_{L,L'}^{IJ} &\equiv \langle JML'S'|\mathcal{M}^I(E_p)|JMLS\rangle \\
&= \sum_{\lambda_i} \sqrt{\frac{2L+1}{2J+1}} \sqrt{\frac{2L'+1}{2J+1}} \langle \lambda_3 \lambda_4|\mathcal{M}^{IJ}(E_p)|\lambda_1 \lambda_2\rangle \\
&\quad \times \langle L'0S'\lambda'|J\lambda'\rangle \langle s_3, \lambda_3, s_4, -\lambda_4|S'\lambda'\rangle \\
&\quad \times \langle L0S\lambda|J\lambda\rangle \langle s_1, \lambda_1, s_2, -\lambda_2|S\lambda\rangle,
\end{aligned} \tag{A1}$$

where $S$ ($S'$) is the total spin of incoming (outgoing) two-particle system and $L$ ($L'$) is the orbital angular momentum of incoming (outgoing) two-particle system. The line 3 and line 4 in Eq. (A1) are C-G coefficients where the notation $\langle j_1, m_1, j_2, m_2|jm\rangle$ is taken. The sum over $\lambda_i$ in Eq. (A1) means that all the helicity combinations of all particles involved should be summed.

When calculating the partial wave amplitudes involved in this paper, some symmetry conditions can simplify the problem [32]. Firstly, $P$-parity conservation demands that,

$$\langle\lambda_3\lambda_4|\mathcal{M}^{IJ}(E_p)|\lambda_1\lambda_2\rangle = \langle-\lambda_3,-\lambda_4|\mathcal{M}^{IJ}(E_p)|-\lambda_1,-\lambda_2\rangle. \tag{A2}$$

Secondly, time-reversal symmetry demands,

$$\langle\lambda_3\lambda_4|\mathcal{M}^{IJ}(E_p)|\lambda_1\lambda_2\rangle = \langle\lambda_1,\lambda_2|\mathcal{M}^{IJ}(E_p)|\lambda_3,\lambda_4\rangle. \tag{A3}$$

And then, if the total spin of the system is conserved, there is another condition,

$$\langle\lambda_3\lambda_4|\mathcal{M}^{IJ}(E_p)|\lambda_1\lambda_2\rangle = \langle\lambda_4,\lambda_3|\mathcal{M}^{IJ}(E_p)|\lambda_2,\lambda_1\rangle. \tag{A4}$$

Considering all the symmetry conditions, the partial wave amplitudes involved can be easily worked out.

For the process $\bar{D}K \to \bar{D}K$, all the particles have zero spin, i.e. $\lambda_i = 0$, $i = 1,\ldots,4$. So, noting that,

$$\langle 00|\mathcal{M}^{IJ}(E_p)|00\rangle \equiv \mathcal{M}^{IJ}(E_p), \tag{A5}$$

then according to Eq. (A1), it gives,
for $J = 0, L = L' = 0$:

$$(\mathcal{T}^{I0}_{0,0})_{\bar{D}K} = \mathcal{M}^{I0}, \tag{A6}$$

and for $J = 1, L = L' = 1$:

$$(\mathcal{T}^{I1}_{1,1})_{\bar{D}K} = \mathcal{M}^{I1}. \tag{A7}$$

For the process $\bar{D}_1 K \to \bar{D}_1 K$, the initial and final state both have three helicity values, so there are nine helicity amplitudes. But after considering the symmetry relations in Eq. (A2, A3, A4), there are only four independent helicity amplitudes. For example, if one choose

$$\begin{aligned}\mathcal{M}^{IJ}_{00} &\equiv \langle 00|\mathcal{M}^{IJ}(E_p)|00\rangle,\\ \mathcal{M}^{IJ}_{+0} &\equiv \langle 00|\mathcal{M}^{IJ}(E_p)|+0\rangle,\\ \mathcal{M}^{IJ}_{++} &\equiv \langle +0|\mathcal{M}^{IJ}(E_p)|+0\rangle,\\ \mathcal{M}^{IJ}_{+-} &\equiv \langle -0|\mathcal{M}^{IJ}(E_p)|+0\rangle,\end{aligned} \tag{A8}$$

as independent amplitudes, and other helicity amplitudes can be deduced from them. Similarly, in this situation, for $L = L' = 0, J = 1$, it can be derived that,

$$(\mathcal{T}^{I1}_{0,0})_{\bar{D}_1 K} = \frac{1}{3}\left\{\mathcal{M}^{I1}_{00} + 4\mathcal{M}^{I1}_{+0} + 2\mathcal{M}^{I1}_{++} + 2\mathcal{M}^{I1}_{+-}\right\}. \tag{A9}$$

for $L = L' = 1, J = 0$,

$$(\mathcal{T}^{I0}_{1,1})_{\bar{D}_1 K} = \mathcal{M}^{I0}_{00}. \tag{A10}$$

and for $L = L' = 1, J = 1$,

$$(\mathcal{T}^{I1}_{1,1})_{\bar{D}_1 K} = \mathcal{M}^{I1}_{++} - \mathcal{M}^{I1}_{+-}. \tag{A11}$$

Note that because of $P$-parity conservation, the difference of orbital angular momentum quantum number between initial and final state is an even number. So, there are amplitudes where $L = 0$, $L' = 2$ in principle which is called S-D mixing [32]. But in this paper, the contribution from $D$-wave is ignored, because it has little effect near the threshold.

For the process $\bar{D}_1 K \to \bar{D}K$, there are only two independent helicity amplitudes which can be denoted as,

$$\begin{aligned}(\mathcal{M}^{IJ}_0)_{\text{trans}} &\equiv \langle 00|\mathcal{M}^{IJ}(E_p)|00\rangle_{\text{trans}},\\ (\mathcal{M}^{IJ}_+)_{\text{trans}} &\equiv \langle 00|\mathcal{M}^{IJ}(E_p)|+0\rangle_{\text{trans}},\end{aligned} \tag{A12}$$

where the subscript "trans" means the transformation process $\bar{D}_1 K \to \bar{D}K$. Then repeating the procedure in Eq. (A1), it can be derived that for $J = 0, L' = 0, L = 1$:

$$(\mathcal{T}^{I0}_{1,0})_{\text{trans}} = -(\mathcal{M}^{I0}_0)_{\text{trans}}, \tag{A13}$$

and for $J = 1, L' = 1, L = 0$:

$$(\mathcal{T}^{I1}_{0,1})_{\text{trans}} = \frac{1}{\sqrt{3}}\left\{\mathcal{M}^{I1}_0 + 2\mathcal{M}^{I1}_+\right\}_{\text{trans}}. \tag{A14}$$

## Appendix B: Analyses for the triangle diagram singularity

The triangle singularity will be discussed here with similar methods in Ref. [33]. The origin of triangle singularity is the three-point loop integral,

$$\begin{aligned}&C_0(p_1^2, p_2^2, p_3^2; m_2^2, m_1^2, m_3^2)\\ &= \frac{\mu^{4-D}}{i}\int\frac{d^D k}{(2\pi)^D}\left\{\frac{1}{[k^2 - m_1^2 + i\varepsilon][(k-p_1)^2 - m_2^2 + i\varepsilon]}\right.\\ &\quad\left.\times\frac{1}{[(k-p_2)^2 - m_3^2 + i\varepsilon]}\right\},\end{aligned} \tag{B1}$$

where $p_3^2 = (p_1 - p_2)^2$. Using standard Feynman parameterization, it can be parameterized as [34],

$$\begin{aligned}&C_0(p_1^2, p_2^2, p_3^2; m_2^2, m_1^2, m_3^2) =\\ &\frac{\mu^{4-D}}{(4\pi)^{D/2}}\frac{2}{p_3^2}\int_0^1 d\alpha_1\frac{1}{\sqrt{4c-b^2}}\\ &\quad\times\left[\arctan\frac{b}{\sqrt{4c-b^2}} - \arctan\frac{b+2(\alpha_1-1)}{\sqrt{4c-b^2}}\right],\end{aligned} \tag{B2}$$

where,

$$b = 1 + \frac{1}{p_3^2}\left[\alpha_1\left(p_1^2 - p_2^2 - p_3^2\right) + m_3^2 - m_2^2\right],$$
$$c = \frac{1}{p_3^2}\left[m_3^2 + \alpha_1\left(m_1^2 - m_3^2\right) - \alpha_1(1-\alpha_1)p_2^2\right] - i\varepsilon.$$

which can be numerically evaluated. In this problem, the three-point loop integral can be expressed as $C_0(P^2, q_3^2, q_{12}^2; m_{D_1}^2, m^{*2}, m_K^2)$. Let $P^2 = s$, $q_3^2 = s_3$, $q_{12}^2 = s_{12}$ in what follows in this paper. The singularity of the integrand function may appear where $\sqrt{4c - b^2} = 0$. Solving this equation with respect to $\alpha_1$, it gives,

$$\alpha_{1\pm} = -\frac{N(s, s_{12}, s_3) \pm 2\sqrt{s_{12}\vartheta(s, s_{12}, s_3)}}{s^2 - 2s(s_{12} + s_3) + (s_{12} - s_3)^2}, \quad (B3)$$

with,

$$\begin{aligned}
N &= m_{D_1}^2(-s + s_{12} + s_3) + m_K^2(s + s_{12} - s_3) \\
&\quad + s_{12}\left(s - s_{12} + s_3 - 2m^{*2}\right), \\
\vartheta &= -m_{D_1}^2\left\{m_K^2(s - s_{12} + s_3) + m^{*2}(-s + s_{12} + s_3)\right. \\
&\quad \left. + s_3(s + s_{12} - s_3)\right\} + sm_K^4 + m_{D_1}^4 s_3 \\
&\quad - m_K^2\left[m^{*2}(s + s_{12} - s_3) + s(-s + s_{12} + s_3)\right] \\
&\quad + s_{12}\left[-m^{*2}(s - s_{12} + s_3) + ss_3 + m^{*4}\right].
\end{aligned}$$

If $\vartheta \to 0^-$, then $\alpha_{1\pm}$ will pinch the integral path and singularities may arise. Solving the equation $\vartheta = 0$ with respect to $s_{12}$, one has,

$$\begin{aligned}
s_{12\pm} &= \frac{1}{2m^{*2}}\left\{m_K^2\left(s + m^{*2}\right) - \left(m^{*2} - s\right)\left(m^{*2} - s_3\right)\right. \\
&\quad + m_{D_1}^2\left(-m_K^2 + s_3 + m^{*2}\right) \\
&\quad \left. \pm\sqrt{\lambda(\sqrt{s}, m_{D_1}, m^*)\lambda(\sqrt{s_3}, m_D, m^*)}\right\},
\end{aligned} \quad (B4)$$

where

$$\lambda(a, b, c) = [a^2 - (b + c)^2][a^2 - (b - c)^2].$$

But the $s_{12\pm}$ in Eq. (B4) does not always give the anomalous threshold. If the $\alpha_\pm$ corresponding to $s_{12\pm}$ are not in the integral domain $[0, 1]$, then the singularities will not arise in the physical sheet. Only when $\alpha_\pm$ are in the domain $[0, 1]$, the singularities may appear. In this section, $m^*$ is a free parameter in the fit, and $s$, $s_3$ are on the mass shell, taking the value of $m_B^2$, $m_D^2$, respectively. So the moving of anomalous threshold with different $m^*$ should be analyzed. The range of $m^*$ values is constrained by $m_B^2 > (m_{D_1} + m^*)^2$, $m_D^2 < (m^* - m_K)^2$, which means $(m_D + m_K) < m^* < (m_B - m_{D_1})$. Substitute the physical masses, $m_B = 5.279$ GeV, $m_D = 1.870$ GeV, $m_K = 0.498$ GeV, $m_{D_1} = 2.421$ GeV, it can be obtained that $2.368$ GeV $< m^* < 2.858$ GeV. The numerical result shows that the $\alpha_{1\pm}$ corresponding to $s_{12-}$ may be in $[0, 1]$. However, it is too early to say that $s_{12-}$ gives the anomalous threshold, since the behavior of two arctangent functions in Eq. (B2) should also be considered.

When $s_{12} \to s_{12-}$, $\alpha_1 \to \alpha_{1\pm}$, $4c - b^2 \to 0$, noting that $\arctan(-\infty) \to \frac{-\pi}{2}$, $\arctan(+\infty) \to \frac{\pi}{2}$, so the signs of $b$ and $b + 2(\alpha_1 - 1)$ should be investigated. If their signs are the same, the two arctangents have same limitations, so their subtraction is zero and cancels the singularity caused by $1/\sqrt{4c - b^2} \to \infty$. If they have different signs, the limitation of the subtraction is $\pm\pi$, combining with that $1/\sqrt{4c - b^2} \to \infty$, the singularity arises and it illustrates that $s_{12-}$ give the anomalous threshold.

After substituting mass parameters, it is found that when $2.368$ GeV $< m^* < 2.858$ GeV, $b > 0$ and $b + 2(\alpha_1 - 1) < 0$, so

$$\lim_{s_{12} \to s_{12-}} \lim_{\alpha_1 \to \alpha_{1\pm}} \left[\arctan\frac{b}{\sqrt{4c - b^2}} - \arctan\frac{b + 2(\alpha_1 - 1)}{\sqrt{4c - b^2}}\right] \\
= \frac{\pi}{2} - \left(-\frac{\pi}{2}\right) = \pi. \quad (B5)$$

So, $s_{12-}$ does give the anomalous threshold. Using the same method, one can find that $s_{12+}$ does not give the anomalous threshold.

Another source of singularities is from arctangent functions. When $x = \pm i$, $\arctan(x)$ is singular. In Eq. (B2) it means that,

$$\frac{b}{\sqrt{4c - b^2}} = \pm i, \quad \frac{b + 2(\alpha_1 - 1)}{\sqrt{4c - b^2}} = \pm i. \quad (B6)$$

Solve these two equations with respect to $\alpha_1$, one can see that the solutions have nothing to do with the c.m. energy square of final state, $s_{12}$. It can not give the anomalous threshold, i.e. two arctangents have no singularity in this problem. So all the singularities in integral Eq. (B2) are caused by $1/\sqrt{4c - b^2}$.

Since the source of singularity is clear, the variation of anomalous threshold with different $m^*$ can be investigated. When mass parameters are substituted, it can be seen that when $2.368$ GeV $< m^* < 2.540$ GeV, $\alpha_{1\pm}$ are not in the integral domain $[0, 1]$. So, it would not cause the singularity in physical sheet. And when $2.858$ GeV $> m^* > 2.540$ GeV, $\alpha_\pm$ are in the domain $[0, 1]$, then the singularity in physical sheet appears. On the other hand, when $2.368$ GeV $< m^* < 2.540$ GeV, the position of anomalous threshold, $s_{12-}$, is located at the upper half-plane of sheet II above the normal threshold. It moves to the threshold with increasing $m^*$ and arrives at the normal threshold $(m_{D_1} + m_K)^2$ when $m^* = 2.540$ GeV. Then, when $m^*$ continues to increase, $s_{12-}$ bypasses the normal threshold and goes above threshold in the lower half-plane of sheet II. According to the analytic continuation of Riemann sheets, above the threshold, the lower half-plane of sheet II is connected to the upper half-plane of sheet I, which is the physical region. So, the singularity caused by anomalous threshold can be "detected" at this time. The trajectory of $s_{12-}$ with increasing $m^*$ is shown in Fig. 6.

Figure 7 shows lineshapes of squared amplitudes of triangle diagrams with different $m^*$. It is shown that when the anomalous threshold is in the upper half-plane of sheet II and gets close to the normal threshold $(m_{D_1} + m_K)^2$, it can produce a spike closing to the normal threshold in sheet I (as shown in Fig. 7a). So it can be concluded that the anomalous threshold in unphysical region still have effect on the physical region when it is close enough to the normal threshold. When the threshold moves to the lower half-plane of sheet II, which connects to the upper half-plane of sheet I, one can see that the spike becomes sharper than before (as shown in Fig. 7b). This is because at this time, the singularity itself, rather its remaining effect is "detected". But it should be noted that the intermediate particle, $D^*_{sJ}$, is not a stable particle, so a finite width should be given. For example, considering $D^*_{sJ}$ are $D^*_{s1}(2700)$ and $D^*_{s1}(2860)$ as argued in Ref. [4] and substituting the physical masses and widths, it can be seen that when it is $D^*_{s1}(2700)$, the central value of the spike is located at about 2.95 GeV while the spike disappears and the lineshape becomes smooth when it is $D^*_{s1}(2860)$ (as shown in Fig. 7c).

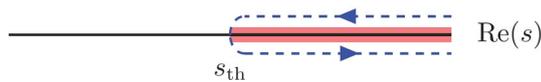

**Fig. 6** Trajectory of $s_{12-}$ on the second sheet with $m^*$ increases from 2.368 GeV to 2.858 GeV

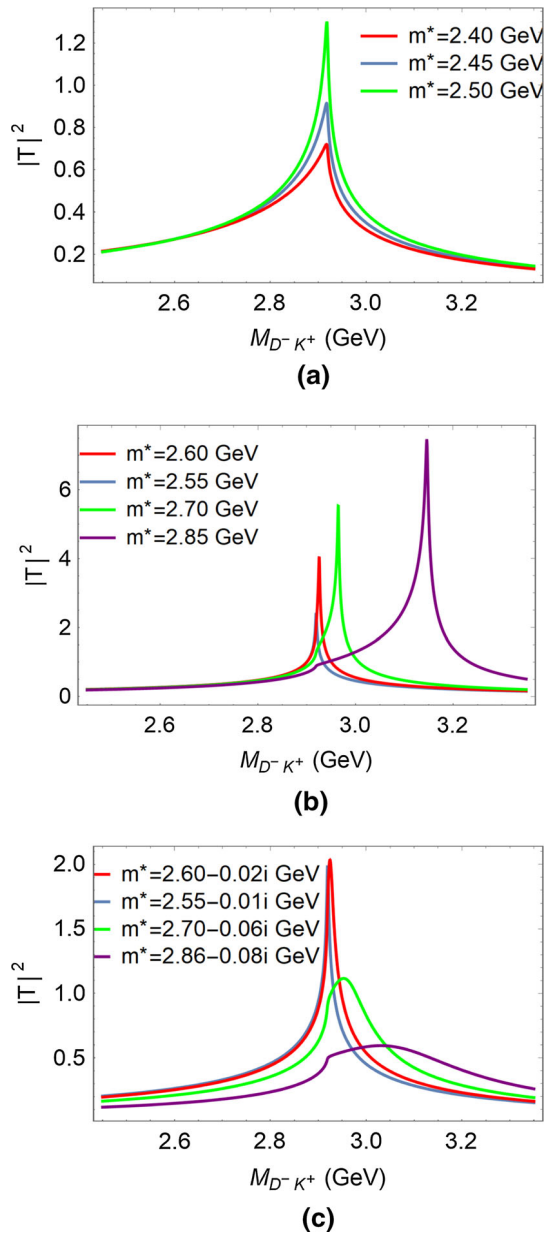

**Fig. 7** Lineshape of triangle diagrams with different $m^*$: **a** $s_{12-}$ located in the upper half-plane of sheet II; **b** $s_{12-}$ located in the lower half-plane of sheet II and $m^*$ has no width; **c** $s_{12-}$ located in the lower half-plane of sheet II and $m^*$ has a finite width